\documentclass[apjl]{emulateapj}
\usepackage{epsfig}

\begin{document}

\title{Turbulence driven by outflow-blown cavities
in the molecular cloud of NGC 1333}

\author{Alice C. Quillen}
\email{aquillen@pas.rochester.edu}
\author{Stephen L. Thorndike}
\author{Andy Cunningham}
\author{Adam Frank}
\author{Robert A. Gutermuth}
\author{Eric G. Blackman}
\author{Judith L. Pipher}
\affil{Department of Physics and Astronomy, University of Rochester, Rochester, NY 14627}
\and
\author{Naomi Ridge}
\affil{Harvard-Smithsonian Center for Astrophysics,  60 Garden Street, Cambridge, MA 02138}

\begin{abstract}

Outflows from young stellar objects have been identified as a possible
source of turbulence in molecular clouds.  
To investigate the
relationship between outflows, cloud dynamics and turbulence, we compare
the kinematics of the molecular gas associated with NGC 1333, traced in
$^{13}$CO(1-0), with the distribution of young stellar
objects (YSOs) within.  
We find a velocity dispersion of $\sim 
1-1.6$km/s in $^{13}$CO  that does not significantly vary across the 
cloud, and is uncorrelated with the number of nearby young stellar 
outflows identified from optical and submillimeter observations.
However, from velocity channel maps we 
identify about 20 depressions in the $^{13}$CO 
intensity of scales $\gtrsim 0.1-0.2$pc and velocity widths $1-3$km/s. 
The depressions exhibit limb 
brightened rims in both individual velocity channel maps and position
velocity diagrams, suggesting that they are slowly expanding cavities.
We interpret these depressions to be remnants of  past YSO 
outflow activity: If these cavities are presently empty, they would fill in on 
time scales of $\sim 10^6$ yr. This can exceed the lifetime of a 
YSO outflow phase, or the transit time of the central star through
the cavity, explaining the absence of any clear correlation between the 
cavities and YSO outflows.  We find that the momentum and 
energy deposition associated with the expansion of the cavities 
is sufficient to power the turbulence in the cloud.
In this way we conclude that the cavities are an important intermediate
step between the conversion of YSO outflow energy and
momentum into cloud turbulent motions.

\end{abstract}


\section{Introduction}

Energetic outflows exert a
strong effect on their parent molecular clouds (for a recent
review see \citealt{ballyreview}). Measurements of the total
kinetic energy present in young stellar outflows imply that
outflows and winds associated with young stellar objects contain
sufficient kinetic energy to excite a significant fraction of the
supersonic turbulence present in molecular clouds and/or unbind
and disperse the cloud \citep{bally96b,ballyreview,knee,warin}.

The molecular cloud associated with the reflection nebula NGC 1333
provides a good setting to study the relation between young
stellar outflows and their influence on parent clouds because it
is an active site of low and intermediate mass star formation. NGC
1333 contains two young star clusters identified in near-infrared
studies \citep{aspin94, lada96}, many Herbig-Haro (HH) objects and
associated molecular outflows (e.g.,
\citealt{bally96a,aspin,knee}) as well as younger, embedded IRAS
sources \citep{sandell,jennings,rodriguez}.

Previous studies have suggested that the outflows present in NGC
1333 could have perturbed the associated molecular cloud. Based on
his study of the star clusters, \citet{aspin} suggested that the
large number of active molecular outflows in the southern region
of NGC 1333 could provide a mechanism for exciting cloud turbulence
and causing subsequent star formation. Based on a study of the
cloud in $^{12}$CO(3-2), \citet{knee} found that molecular
outflows observed in the cloud could provide enough kinetic energy
to accelerate the entire cloud by a few km/s. Dust ridges and
shells seen in submillimeter continuum can be associated with
outflows implying that the cloud itself has been considerably
modified and, perhaps, disrupted by the outflows
\citep{sandell,warin,lefloch,bally96a}.

Many authors have argued that outflows could stimulate star
formation or be responsible for turbulence in molecular clouds.
These arguments usually rely on estimates of the energy present in outflows
compared to that associated with turbulence in the clouds \citep{ballyreview}.
Studies of individual objects,
however, express a more complicated picture.
In particular the
explicit nature of the coupling between cloud material and
outflows could be better explored.
Molecular outflows are identified through millimeter radio
emission from molecules such as CO extending beyond the more collimated
emission seen in HH objects in the visible (e.g., \citealt{bence96,yu99}).  In giant outflows, CO emission
associated with the outflow can extend to a few arcminutes from the
central source \citep{chernin95,cernicharo}. Unified
models suggest that it is the narrow bipolar jet with, perhaps,
a wide-angle wind component, that entrains molecular material
which is then seen as the CO molecular outflow
\citep{raga93,chernin94,delamarter00,lee02,arce2004,gardiner03}.
We note that high angular resolution observational studies
have revealed evidence for wide angle winds in addition to
collimated jets (e.g., \citealt{arce2002a}).

In $^{12}$CO lines such as the  $J=$1-0 transition
the optical depth increases rapidly near the velocity centroid of a molecular
cloud.
However, the optical depth decreases with increasing velocity offset and
so $^{12}$CO can be used to trace lower density material associated with
outflows (e.g., the survey of NGC 1333 by \citealt{knee}).
However to trace the structure at higher densities,
a more optically thin tracer such as $^{13}$CO or C$^{18}$O is needed.
\citet{ridge03} have carried out a survey of nearby molecular clouds in
these two emission lines, providing a database with which to investigate
the mechanism and timescale for the dissipation of gas around individual
stars and  clusters.    NGC 1333 has been previously surveyed
in the same molecules by \citet{warin}.
The overall structure of the warm gas in the cloud 
is expected to be well traced by  the C$^{18}$O emission, with depletion
or freeze-out
of CO only occurring in small dense pockets \citep{bergin}.

Outflows and winds associated with young
stellar objects produce high velocity gas, but also can evacuate
regions in their host molecular cloud.
Cavities associated with outflows have been seen previously, particularly
in low opacity tracers such as $^{13}$CO 
(e.g., \citealt{warin,arce2002b,lefloch,welch}).
In the Circinus outflow, \citet{bally99} described dusty filaments confining
1pc sized cavities that were also seen in their $^{13}$CO and C$^{18}$O maps.
They suggested these cavities were fossil remnants of previous outflows.
\citet{snell80} described a 15 km/s velocity, pc sized double lobed cavity
associated with L1551.
\citet{reipurth98} described an evacuated pc long chimney associated
with HH-310.
With higher resolution CO interferometry, \citet{lee02} described
complex outflow morphology such as shell-like and multipolar structures,
multiple cavities, and asymmetric lobes.
\citet{arce2004} showed that the outflow of L1228 had evacuated a
conical shaped cavity, and had morphology and kinematics
consistent with entrainment of gas by both collimated and wide angle
outflow components.  \citet{welch} showed that a 
0.05pc expanding shell in Taurus could have been caused
by outflows driven by the binary XZ Taurus.
In the NGC 1333 cloud,
\citet{warin} suggest that the SVS13, IRAS 2 and IRAS 4 outflows
have created the cavity that lies between SVS13 and HH12.  However
\citet{knee} found that the molecular outflow associated with IRAS 4 did not
overlap with this cavity.

In a few cases, theoretical models for expanding spherical cavities
{\it i.e.} wind blown bubbles, have been applied to cavities
opened by outflows. \citet{koo92a} applied their wind blown bubble
model to the HH 7-11 region, located in the NGC 1333 cloud. In the
context of a slow wind, \citet{koo92a} suggested that cavities
opened in molecular clouds by outflows were consistent with their
estimated mechanical luminosities.
\citet{raga04} modeled the limb-brightened cavity associated with HH 46/47
\citep{noriega} as a bow shock driven by a perfectly collimated jet.

Most observational studies have suggested that existing HH objects
and outflows could be responsible for creating cavities (e.g.,
\citealt{knee, koo92a, warin, lefloch,lee02,arce2002b, noriega, raga04}), 
and have identified the most prominent outflows evident in the winds of the
CO channel maps or seen as HH objects with the most prominent
cavities. Since the duration of the strong outflow period of a
protostar may be short compared with the life of a young cluster,
the history of the coupling between the outflow and 
molecular cloud must be considered.
Indeed, some studies have also suggested that cavities in
molecular clouds could be tracers of previously active outflows
(e.g., \citealt{bally99}). In the vicinity of SVS13 in NGC 1333,
both possibilities have been suggested \citep{warin,lefloch}.
\citet{warin} suggested that the star cluster north of SVS13 was
associated with a cavity, whereas \citet{lefloch} identified two
cavities south of SVS13 and associated one them with a cone shaped
outflow emanating from the infrared source SVS13.


To investigate the coupling between outflows and the dynamics of
molecular clouds, we compare the spatial distribution of outflows
in NGC 1333 with the properties of its molecular cloud.
Specifically we compare the lower optical depth $^{13}$CO
millimeter velocity cubes obtained by \citet{ridge03} with near
infrared sources and the locations of outflows identified by previous surveys.
In addition we compare structures seen in the
molecular gas distribution to the locations of young stellar
objects.

We adopt a distance to NGC 1333 of 212 pc based on the distance
estimated to the star BD$+30^\circ$549 that illuminates the nebula
\citep{cernis90}.
This distance is nearer than the estimate 318 pc that is based on
on the {\it Hipparcos} parallactic measurements of the Perseus OB2
association \citep{dezeeuw99}. At the distance of 212 pc, 1'
corresponds to 0.062 pc.

\section{Mean velocity field and velocity dispersion of the molecular gas}

If outflows associated with young stellar objects
are responsible for turbulence excited in the cloud,
then we might expect regions that contain large numbers of outflows
to reside in a more turbulent molecular medium.
In this section we compare the integrated properties 
of the molecular gas,  as traced in $^{13}$CO by \citep{ridge03}
with the distribution of previously identified outflow sources.

Of the NGC 1333 velocity cubes presented and discussed
by \citet{ridge03} the highest signal
to noise data set for NGC 1333 is
the $^{13}$CO(1-0) data cube obtained at the Five College Radio
Astronomy Observatory (FCRAO). Because of
its higher signal to noise we primarily use this
velocity set for our comparison. The angular resolution of these
data is  $47''$ and the spectral resolution corresponds
to a velocity spacing between each channel of 0.133 km/s.
For NGC 1333, $47''$ corresponds to $\sim .05$pc.

\begin{figure*}
\epsscale{1.00}
\plotone{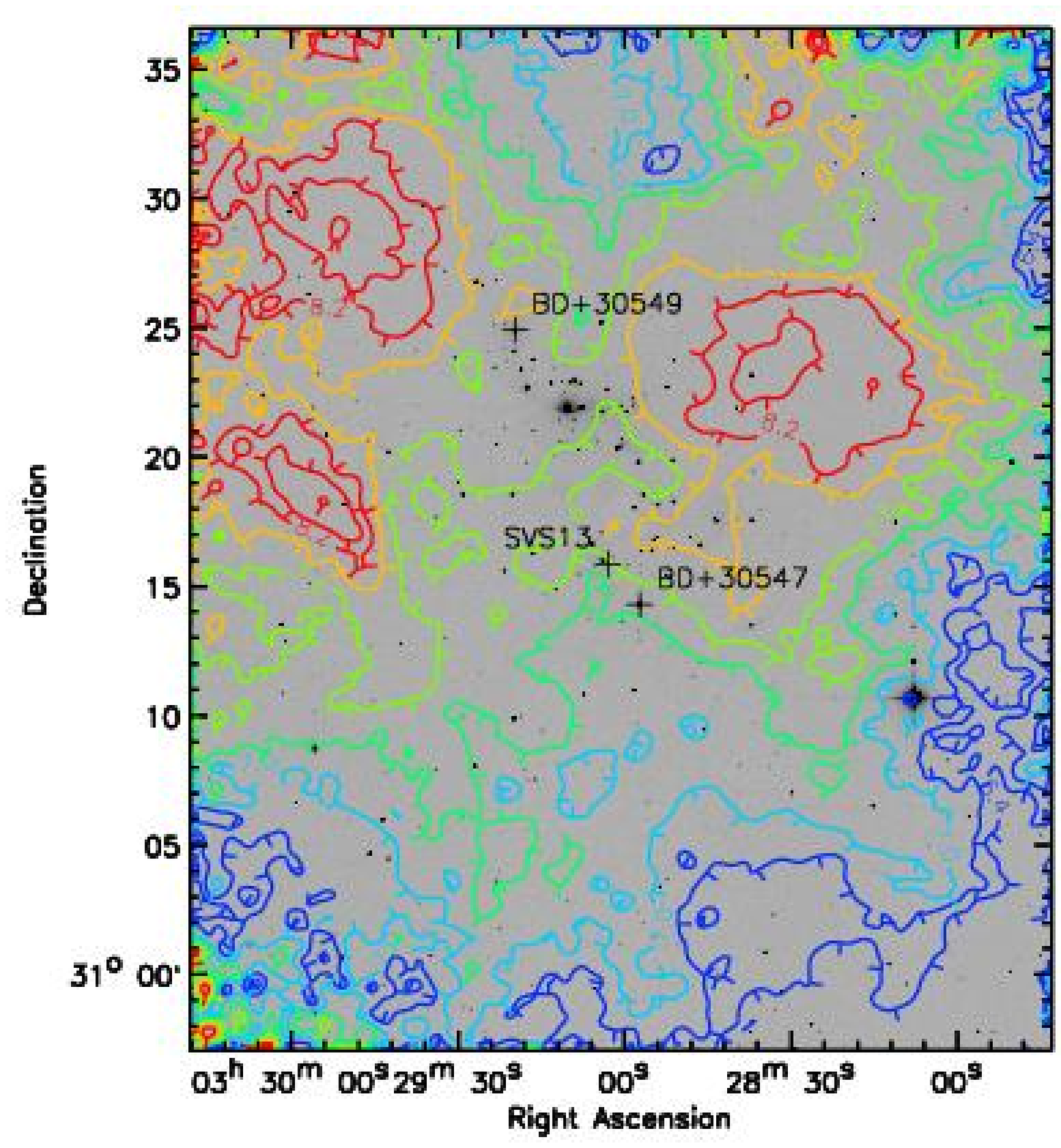}
\figcaption{
Mean velocity of the $^{13}$CO(1-0) velocity cube
contours (see \citealt{ridge03})
overlayed on the Ks band 2MASS
image shown as grayscale.
Contour spacing is 0.3km/s and given with respect to the local standard
of rest.  Red contours are at 8.2 km/s and higher.
The yellow contour is at 7.9 km/s.
Green counters are at 7.3 and 7.6 km/s.
The blue contours are at 7.0 km/s and lower.
The bright region just to the left and below the center of the frame
contains the star SVS13 which has been
used as a reference point by previous works (e.g., \citealt{knee}).
Also shown for reference are the B stars BD+30547
and BD+30549. BD+30549 illuminates the reflection nebula.
The region south of SVS-13
is moving at a higher mean velocity (by about 0.5km/s) than the
northern region.
The angular resolution of the $^{13}$CO data is 47''.
Coordinates are given with respect to epoch J2000.
\label{meanvelocity}
}
\end{figure*}

\begin{figure*}
\epsscale{1.00}
\plotone{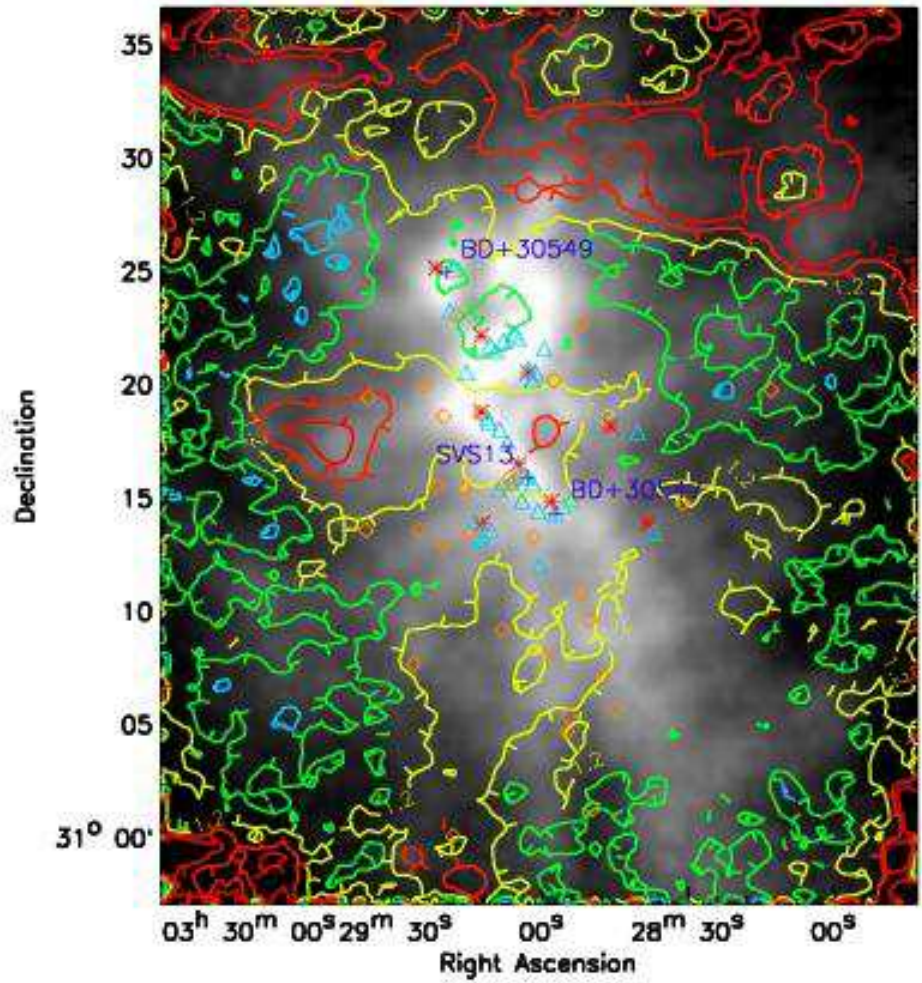}
\figcaption{
Velocity dispersion (rms) about the mean from the $^{13}$CO(1-0) velocity cube
are overlayed on the integrated
$^{13}$CO intensity (grayscale).  Contour spacing is 0.3 (km/s).
The red contours are at 1.5 and 1.8 km/s.  The yellow contour
is at 1.2 km/s. The green contour is at 0.9 km/s.
Much of the structure in the velocity dispersion map is due to
the low signal to noise at the cloud outer edges.
While the cloud shows significant variations in the mean velocity,
little variation in the velocity dispersion is seen across the cloud.
For comparison,
IRAS sources from \citet{jennings} are shown as red stars.  HH objects
from \citet{HHcat} are shown as orange
diamonds. Compact submillimeter sources from \citet{knee}
are shown as blue triangles.
The B stars BD+30547 and BD+30549, and infrared source SVS13
are shown as dark blue crosses.
Two depressions in the molecular gas distribution are seen,  one
just to the north west of SVS13 and the other 7 arcminutes north
and slightly east of SVS13.
These depressions are not coincident with bright
infrared sources, HH objects or early type stars.
\label{mom2}
}
\end{figure*}

In Figure \ref{meanvelocity} we show the mean velocity
measured from the $^{13}$CO(1-0) data cube as contours,
overlaid on the 2MASS Ks band image shown as grayscale.
In Figure \ref{mom2} we show the velocity dispersion measured from
the $^{13}$CO(1-0) data cube, overlaid on the integrated
intensity of the velocity cube. The mean velocity, $<v>$, is the intensity
weighted mean velocity at each position of the map measured from
the entire velocity cube and  
is computed as $<v> = \sum_i v_i I_i/ \sum_i I_i$ 
where $I_i$ and $v_i$
are the velocity and intensities of the individual pixels. 
The velocity dispersion, $\sigma$, is the
square root of the intensity weighted square of the velocity subtracted
by the mean;  $\sigma^2 =\sum (v_i - <v>)^2 I_i /\sum_i I_i$.

We might expect an increase in the velocity dispersion of the molecular
cloud in region that contain many outflow sources
or many embedded young objects. However,
from inspection of Figure \ref{mom2} we find
that the velocity dispersion as measured from $^{13}$CO does not
significantly vary across the molecular cloud.  The velocity dispersion
ranges between 1.0 and 1.6 km/s across the cloud. 
This velocity range is well within the blue-shifted wings 
seen in $^{12}$CO that are 10 km/s below the cloud velocity mean 
from outflows
associated with the most prominent HH objects \citep{knee}.

\begin{figure*}
\epsscale{1.00}
\plotone{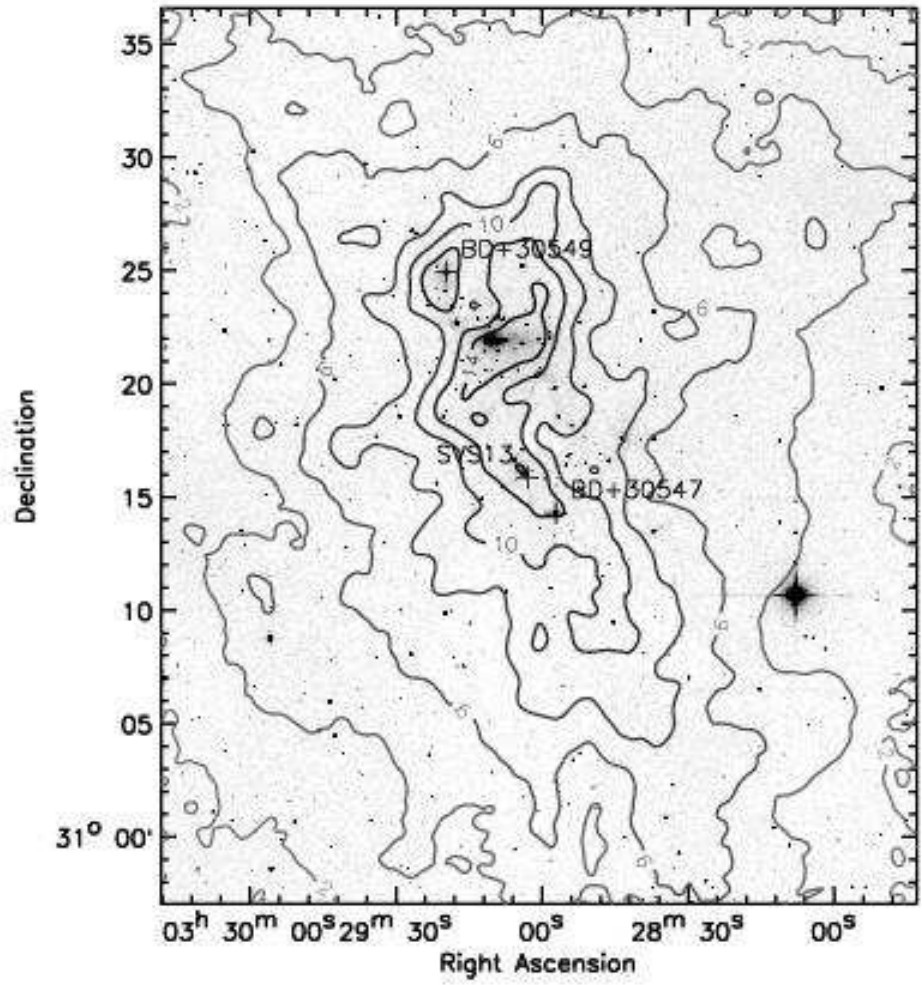}
\figcaption{
Integrated $^{13}$CO intensity (as contours)
overlayed on top of the Ks band 2MASS image (as grayscale).
Contours are shown  2 K km/s apart with the lowest and highest contours
at 2 and 14 K km/s.
The 2MASS image shows
T-Tauri cluster stars as well as background stars.
\label{mom4}
}
\end{figure*}

Hydrostatic equilibrium models for clouds predict relations
between velocity dispersion, density and radius (e.g., \citealt{myers}).
However, Fig. \ref{mom2} shows no strong relation between
integrated intensity and velocity dispersion, 
as would be expected from hydrostatic cloud models.
There is little significant structure in the constant 
dispersion contours, though the dispersion is somewhat higher 
near the lower $^{13}$CO intensity region to the north of SVS-13.
The somewhat higher velocity dispersion in this depression or cavity
might be explained with a scenario in which a cavity is evacuated
by winds driven from sources within the cavity.
Two star clusters have been identified in the near-infrared
imaging by \citet{aspin94}.  
One cluster is located in the vicinity of this cavity and another
is located near the depression in the integrated $^{13}$CO
about 7' to the north of SVS13. These clusters can be seen 
in the K-band stellar distribution in Fig. \ref{mom4}.
The outflow region associated with SVS-13 (the HH7-11 region)
does not have a larger velocity dispersion
than other regions of the cloud.  Likewise the HH4, HH2 and HH12 regions
associated with outflows seen in $^{12}$CO by \citet{knee} also do not stand
out in the velocity dispersion maps.
Significant structure correlated with positions of known outflows
in the cloud random motions as traced by the velocity dispersion
are not observed in the the lower opacity tracer $^{13}$CO.

Outflows previously identified in
$^{12}$CO as high velocity wings (with blue
and red-shifted emission greater than a few km/s from the cloud
mean) by \citep{knee} are not evident in the high velocity wings in
the $^{13}$CO or C$^{18}$O data of \citep{ridge03}. 
At velocities below 5 km/s and above 10 km/s
little emission is detected in the individual $^{13}$CO channel maps 
(see Figure \ref{fig:gray}).
\citet{knee} found outflows associated with HH6, HH12, HH7-11, IRAS 2 
and IRAS 4; their CO maps did not extend all the way to IRAS 1. 
Surveys in $^{12}$CO detect outflows whereas those in $^{13}$CO often do not
(e.g., \citealt{warin}). This is most likely because the $^{12}$CO 
emission is more
sensitive to optically thin and low density regions, but less
capable of penetrating denser higher opacity regions (e.g., \citealt{arce2004}).
A deeper, more sensitive velocity cube in $^{13}$CO 
might reveal additional structure near the individual
outflows, particularly at velocities more than 3 km/s above or
below the mean cloud velocity.
Outflows traced in the high velocity wings of molecules such as $^{12}$CO
are associated with a low density
medium that is faint and so difficult to detect in $^{13}$CO.

Because we see little high velocity gas in $^{13}$CO and no large 
spatial variations in the velocity dispersion,
we find little evidence for turbulence
in the denser regions of the molecular cloud that has been directly induced
by the high concentration of high velocity outflows discovered at
other wavelengths.   
We conclude that the bulk of the cloud, which we expect is
traced in $^{13}$CO, does not contain much high velocity gas 
associated with active outflows.

In contrast to the velocity dispersion map, 
there is quite a bit structure in the mean velocity
field shown in Figure \ref{meanvelocity}. The mean velocity field
exhibits a gradient with velocity increasing from the south to the
north of the cloud by about a km/s. There is also a jump in
velocity of $\sim 0.3$ km/s separating the southern region in NGC
1333S from the rest of the cluster.


\section{Cavities}

In the previous section we found that 
the integrated properties of the $^{13}$CO velocity cube were
not strongly influenced by outflows.
In this section we inspect individual
channel maps and position velocity diagrams from the $^{13}$CO velocity
cube to better investigate the possibility that
outflows have affected the structure of the molecular cloud.

\begin{figure*}
\epsscale{1.20}
\plotone{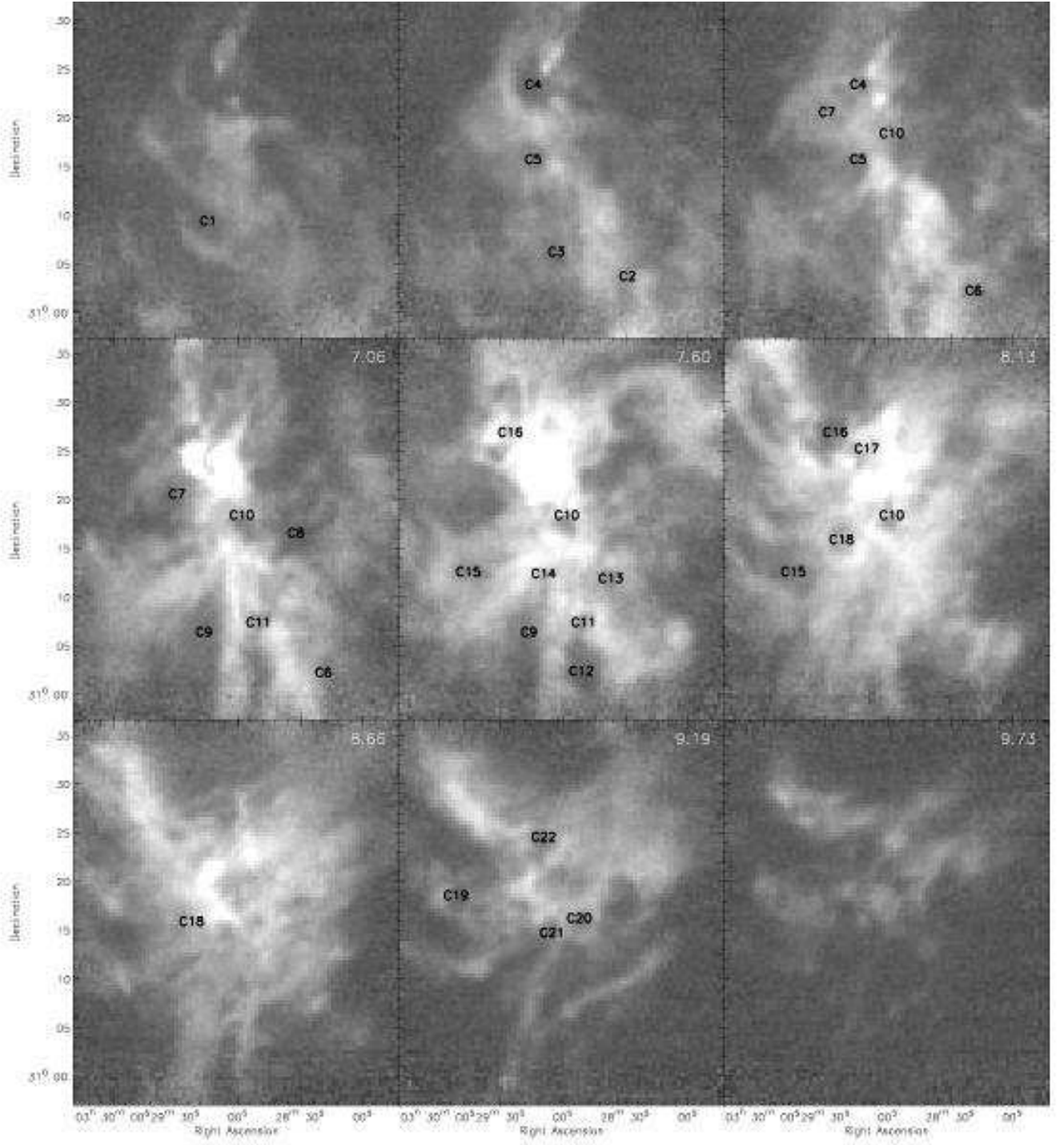}
\figcaption{
Channel maps are shown for a subsample of the velocity cube.
The velocity spacing between the channel maps displayed is 4 channels
or 0.53 km/s.
Each channel has a velocity width of 0.133 km/s.
The velocity of the channel shown in each panel is shown on the top right
hand side of the image.
Multiple shells and cavities are evident in the individual channel
maps.
Depressions in the intensity
identified in both position velocity diagrams and individual
channel maps are labeled here on the channel maps.
The properties of these cavities are listed in Table 1.
\label{fig:gray}
}
\end{figure*}

Instead of high velocity gas associated
with outflows, we see shells and cavities or depressions
in the gas distribution in individual channel maps; 
see Figure \ref{fig:gray}.    
Cavities in the $^{13}$CO integrated
intensity have been described by previous studies
(e.g., \citealt{warin,lefloch}).
Because these depressions in intensity 
are seen in more than one neighboring velocity channel,
they are likely to be real and not a result of noise in the data.
Different cavities appear at different velocities, suggesting
that they are often not directly related to one another.
Some cavities can disappear at velocities higher or lower than the central value
where the depression is most prominent, although CO emission can remain strong.
This suggests that the cavities can lie within the cloud,
and are not exclusively a result of filamentary structure
near the cloud edges.  
Previous studies comparing maps of line emission in C$^{13}$O,
$^{13}$CO, NH$_3$ and CS have found that depressions in the line
intensity in $^{13}$CO and C$^{18}$O are primarily due to
lower gas density rather than depletion of CO, temperature variations 
or high opacity  \citep{warin,lefloch}.
From inspection of the channel maps, we find that
the most prominent cavities are also evident in the 
lower opacity and lower signal to noise 
C$^{18}$O velocity cube of NGC 1333 
(for additional discussion
on morphology in C$^{18}$O compared to that seen in $^{13}$CO
see section 3.1 below).

The velocity gradient across the NGC 1333 molecular cloud
is fairly smooth or shallow (Fig. \ref{meanvelocity}), so evacuated
or lower density
regions stand out as depressions in individual channel maps.
The rims of many of these evacuated regions
are oval in the low angular resolution
$^{13}$CO images (see Fig. \ref{fig:gray}).  If molecular gas
has been evacuated from a central region then we would expect that
the resulting cavity would appear limb brightened.  This appears
to be true for a number of cavities evident in the individual channel maps
(e.g., C1, C4).

In Figure \ref{fig:gray} we have identified and labeled 
a number of cavities.
Their estimated central positions and velocities are listed in Table 1
in order of increasing central velocity.
These cavities correspond to depressions in channel maps
that we have also identified as depressions in position velocity diagrams.  
To list a cavity, we required that they be visible (by eye)
as depressions in the intensity seen in 
both channel maps and position velocity diagrams, however
we did not require a specific contrast level in the intensity.
In some cases the intensity of the edge of
the cavity is only $\sim 50\%$ that of the center in a particular
channel,
in other cases the intensity inside the cavity is a few times lower
than that of its rim (e.g., C10).
We also required that the depression be surrounded on at least three sides
by higher intensity emission in a channel map.

In Fig. \ref{fig:pv} we show position velocity
diagrams extracted from the $^{13}$CO velocity cube in narrow horizontal strips.
The width of the strip corresponds to one pixel in the velocity cube or $25''$
which is less than half the beam width.
The $x-$axes of these plots is the RA(J2000) 
and the $y-$axis, the velocity in km/s.
Figure \ref{fig:pv} shows position velocity diagrams extracted 
for a number of different horizontal
(East-West strips) each separated by $2'$ in declination.
The DEC(J2000) of each strip is shown in the upper right hand 
side of each subplot.
Figure \ref{fig:pvv} shows a similar set of position velocity diagrams,
however each plot is extracted along a narrow strip oriented vertically.
Consequently the $x$-axes are DEC(J2000). The separation in RA between strips
is $2'$.  
The RA(J2000) of each strip is shown on the upper right hand side of each
subplot.

\subsection{Cavity Properties}

\begin{figure*}
\epsscale{0.87}
\plotone{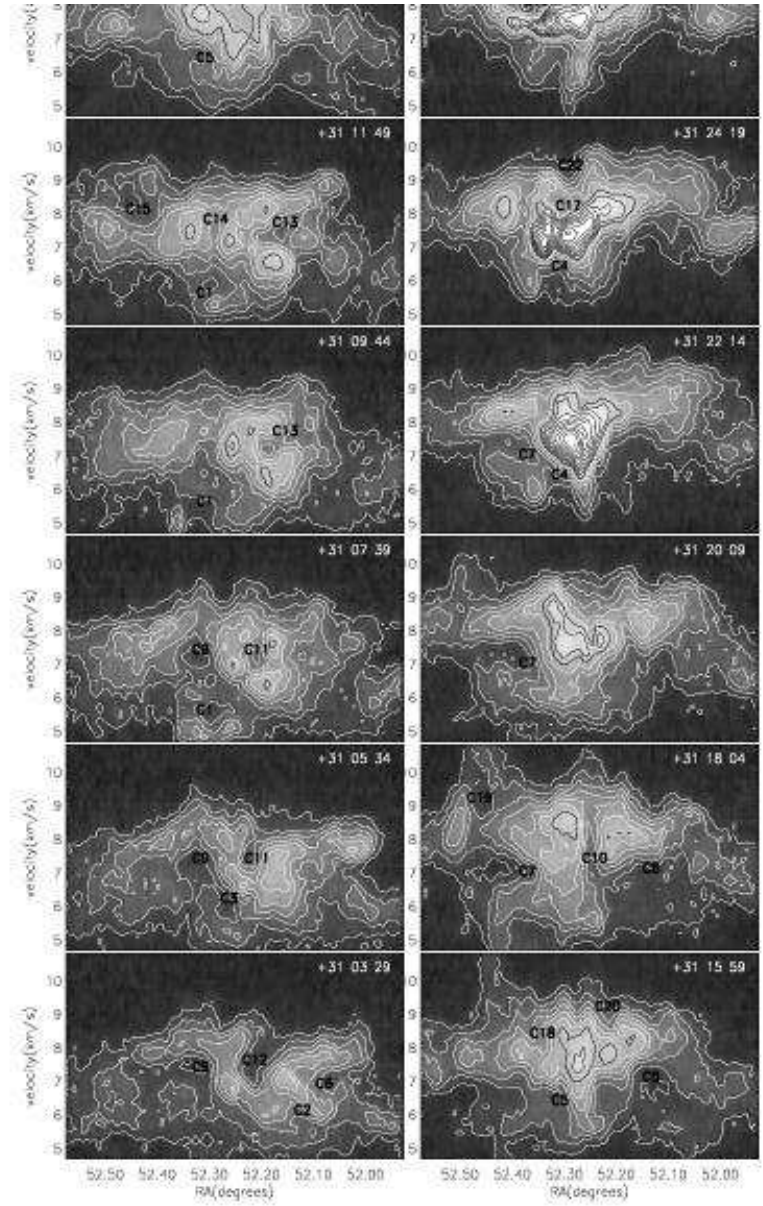}
\figcaption{
Position velocity maps.  Each map is located at a different
declination in the $^{13}$CO(1-0) velocity cube. 
This DEC is shown on the top right hand
side of each panel. Velocity is shown as the vertical
axis, whereas the horizontal axes are RA(J2000).
The contour spacing is 0.5K with the lowest contour at 0.5K.
Depressions in the intensity
identified in both position velocity diagrams and individual
channel maps are labeled here on the channel maps.
The properties of these cavities are listed in Table 1.
\label{fig:pv}
}
\end{figure*}

\begin{figure*}
\epsscale{0.87}
\plotone{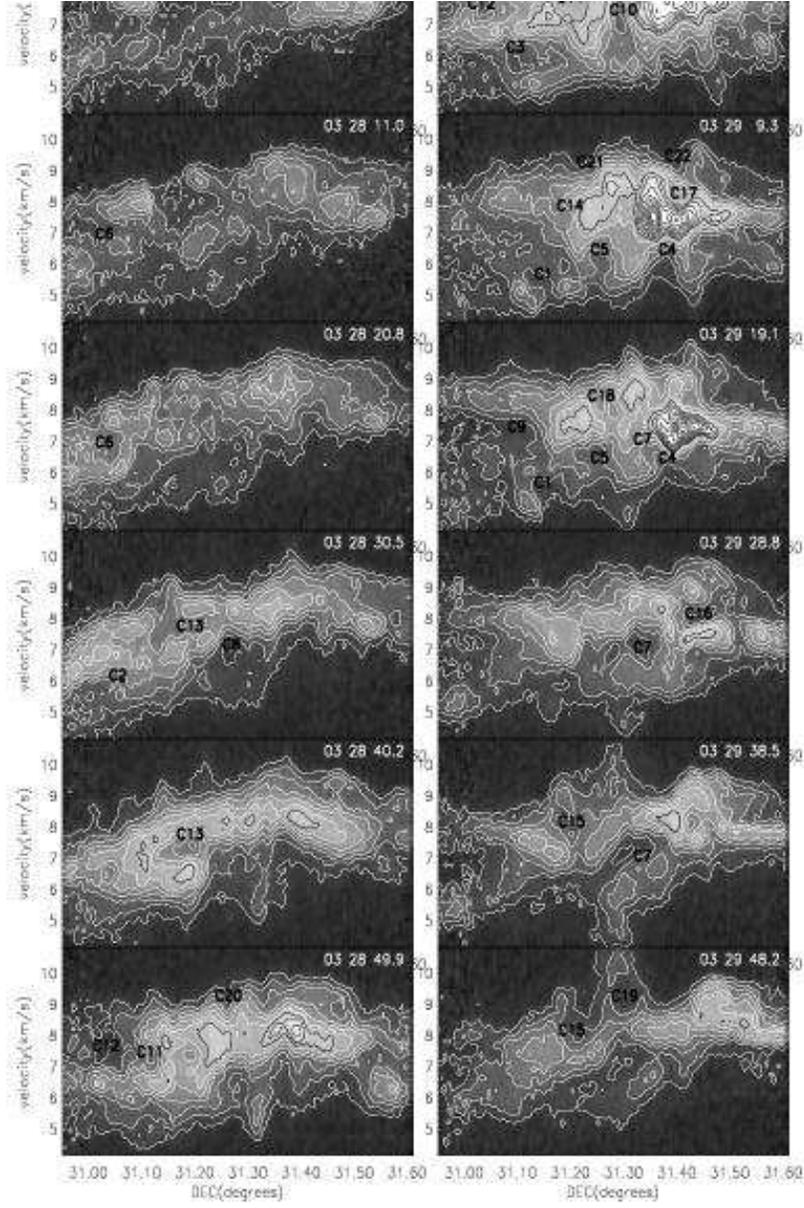}
\figcaption{
Position velocity maps.  Each map is located at a different
right ascension in the $^{13}$CO(1-0) velocity cube. 
This RA is shown on the top right
hand side of each panel. Velocity is shown as the vertical
axis, whereas the horizontal axes are DEC(J2000).
The contour spacing is 0.5K with the lowest contour at 0.5K.
\label{fig:pvv}
}
\end{figure*}

The cavities that we have identified in Figures
\ref{fig:gray}, \ref{fig:pv} and \ref{fig:pvv}
have typical widths of 2--$4'$ corresponding to 0.1--0.2pc.
The density contrast suggested by the difference between the emission
intensity at the center of the cavity compared to that in the edge brightened
rims is a factor of 2--4.  Cavities are evident in channel maps and are
also seen as depressions in the position velocity diagrams.  
For many cavities, at the cavity center 
emission can be seen red-shifted from the mean central
velocity and also blue-shifted from this mean. 
If the cavity is expanding, the red-shifted emission we would
interpret to be behind the cavity moving away from us and the blue-shifted
would correspond to material in front of the cavity moving toward us.
The width in velocity space of these cavities is approximately 1 km/s except
for extreme cavities such as C10 which is seen in the channel maps
between 6 and 9 km/s.  Some cavities are bounded by both red-shifted
and blue-shifted emission (e.g., C11) others appear to have one open end
(e.g., C10).

Many cavities are connected to or associated with other cavities.  
In other words they are near in position and velocity to other cavities. 
For example, C11, C12, and C13
together may form one long tubular region oriented north-south
in the cloud.  C9 and C14 are a smoothly connected triangular region, and
C7 and C4 could be connected.  C15 and C5 could be a connected tubular
region running south-east.
In the southern region of the cloud, the interconnected
cavities seem to run north-south, whereas on the eastern side of the cloud,
cavities tend to be oriented east-west.
The orientation of the cavities would be consistent with a scenario
where the cavities were associated with previous or relic
outflows driven from sources near the center of the cloud.
The center of the cloud
contains two clusters of young stars which
are evident in near-infrared images (see Figure \ref{mom4}).
These sources could have driven active outflows a few hundred
thousand to about a million years ago \citep{aspin}.

Cavities that have red-shifted and blue-shifted rims (e.g, C13,
C11, C9, C3, C9, C8, C1, C7, C16) are most easily
interpreted in terms of expansion.
The radial expansion rate is slow,  half to 1 km/s typically, 
however this is high enough
that is above the sound speed for the cold temperatures typical of molecular
clouds $\sim 0.3$ km/s, though it is similar to the cloud's velocity
dispersion.
A few of the cavities (e.g., C5, C10 and C19) have larger
radial expansion velocities (1.2-2.5 km/s), 
above the velocity of turbulent motions in the cloud.
The red-shifted and blue-shifted rims would be consistent with
a cavity that is slowly expanding in the cloud.
Because emission can be brighter in the red-shifted or blue-shifted
rims, compared to nearby material (e.g., C9, C16), 
the cavities are likely to
be expanding rather than boundaries of
regions of the cloud which have only been evacuated by outflows.

An expanding spherical shell would appear as an oval symmetrical
about the spatial axis in
position velocity diagrams. If the position velocity
diagram is extracted along the major axis of an expanding
ovoid shell, the cavity would appear as
a skewed oval (one oriented along a diagonal line)
in the position velocity diagram,
with one side red-shifted and the other the other side blue-shifted
(e.g., \citealt{welch}).
If the position velocity diagram is extracted along a different axis,
the cavity would also appear as an oval in the position  velocity diagram
but with central position and velocity dependent upon the location
of the extracted position velocity diagram.
Since we were searching the entire cloud
for cavities, we have extracted in Figures \ref{fig:pv}, \ref{fig:pvv}
position velocity diagrams in north-south and east-west strips in the cloud.
If non-spherical (elongated or ovate)
cavities are oriented at different positions with respect
to the line of sight and the angle that the position
velocity diagrams are extracted, we would we expect
to see skewed ovals in the position velocity diagrams
as well as ovals with positions that vary with position
in neighboring position velocity diagrams (e.g., see \citealt{welch}).

In the position velocity diagrams we see examples of cavities
which could be interpreted as expanding non-spherical
or ovate shells at different orientations.
For example, 
C10 appears to be oriented nearly along the line of sight.
It is a small cavity that spans 3 km/s and has almost no
skew in the position velocity diagrams. In other words,
the cavity central position is not strongly dependent upon
the velocity channel, and the central velocity does
not strongly depend on position on the sky.
However, C18 and C19 together may be an example of a cavity
that is skewed in the position velocity
diagram, with the western side red-shifted and the eastern side
blue-shifted.  The connected cavities (e.g., C11, C12, C13)
could be examples of cavities which change
location in both velocity and position in
different position velocity diagrams.
It is likely that we are seeing expanding ovoid
cavities oriented at different
angles with respect to the viewer as well as those oriented
along the line of sight, such as C10.
C10 is seen at almost all velocities (except near 5 km/s) suggesting
that it passes through the entire cloud. 
Other cavities, such as C12 and C15
are seen only at higher or lower velocities, suggesting that they pass through
only one side of the cloud.

A few cavities exhibit high velocity wings (of order 2 km/s from
the mean which is above the velocity dispersion of the cloud).
C10 is extraordinarily broad, extending through a number of channel
maps. The cavity has a width of about 3km/s.
C5 could be connected to blue-shifted high velocity gas (at 4 km/s),
see the panel at RA=3h29m19s(2000)
in Fig. \ref{fig:pvv}. The evacuated triangular shape in the position
velocity diagram suggests that a conical shaped cavity could have been
opened.  This shape has been seen in previous works.  For example,
\citet{arce2004} suggested that IRAS20582+7724
exhibited a cavity which had been opened by a wide angle wind.
C19 is also a triangular shaped cavity in the position velocity diagram
(see the panel at DEC$=+31^\circ18'04''$(2000) 
in Fig. \ref{fig:pv}) and is associated
with red-shifted gas 2km/s above the mean of the cloud.
Because they are associated with higher velocity emission on their rims,
C10, C5 and C19,  could be examples of younger cavities.

\begin{figure*}
\epsscale{0.87}
\plotone{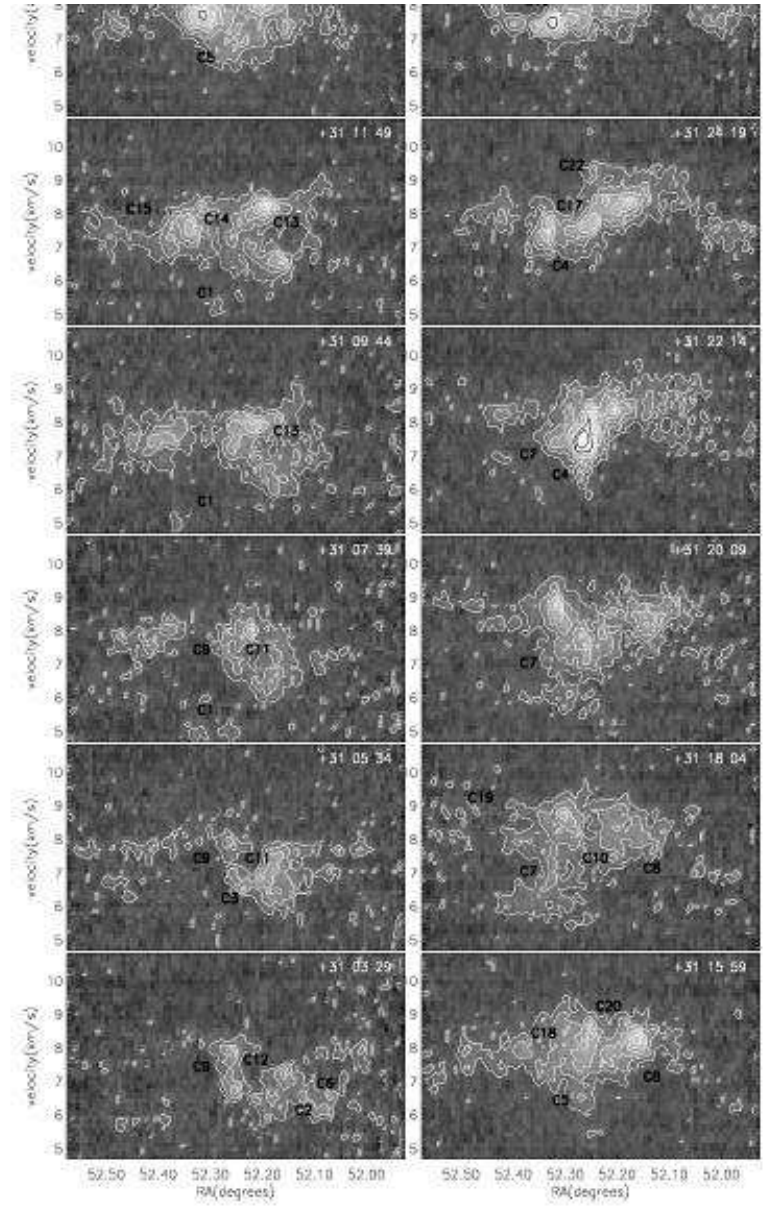}
\figcaption{
Position velocity maps from the C$^{18}$O(1-0) velocity
cube. Similar to Figure \ref{fig:pv}. 
The contour spacing is 0.2K with the lowest contour at 0.2K.
Cavities in the denser regions of the cloud identified from 
the $^{13}$CO(1-0) velocity cube are also evident in the 
C$^{18}$O position velocity maps.
\label{fig:pve}
}
\end{figure*}

The C$^{18}$O(1-0) velocity cube has lower signal to noise than
the $^{13}$CO(1-0) cube, however the 
C$^{18}$O(1-0) line also has lower optical depth.
Features detected in $^{13}$CO in the outer more tenuous regions 
are more difficult to detect in the C$^{18}$O velocity cube.
However, cavities identified in the denser regions of the cloud
should be visible in both velocity cubes.
In Figure \ref{fig:pve} we show position velocity plots
of the C$^{18}$O emission extracted
in the same regions as those of Figure \ref{fig:pv} which displays the 
$^{13}$CO(1-0) emission.
From a comparison between Figure \ref{fig:pv} and \ref{fig:pve}
we see that cavities such as C6, C7, C11, C12, C13, C14 and C16
which have higher intensity rims are seen in both velocity cubes.
The cavities which are not detected in the C$^{18}$O velocity cube
lie in the more diffuse regions of the cloud.  Particularly prominent
in the C$^{18}$O position velocity plot is the cavity C16.
The good correspondence between the morphology seen in these two
figures confirms that the cavities identified in $^{13}$CO
in the denser regions of the cloud are not due to noise in the data,
and are not artifacts introduced by variations in optical depth.

\subsection{The relation between cavities and infrared sources}

\begin{figure*}
\epsscale{1.20}
\plotone{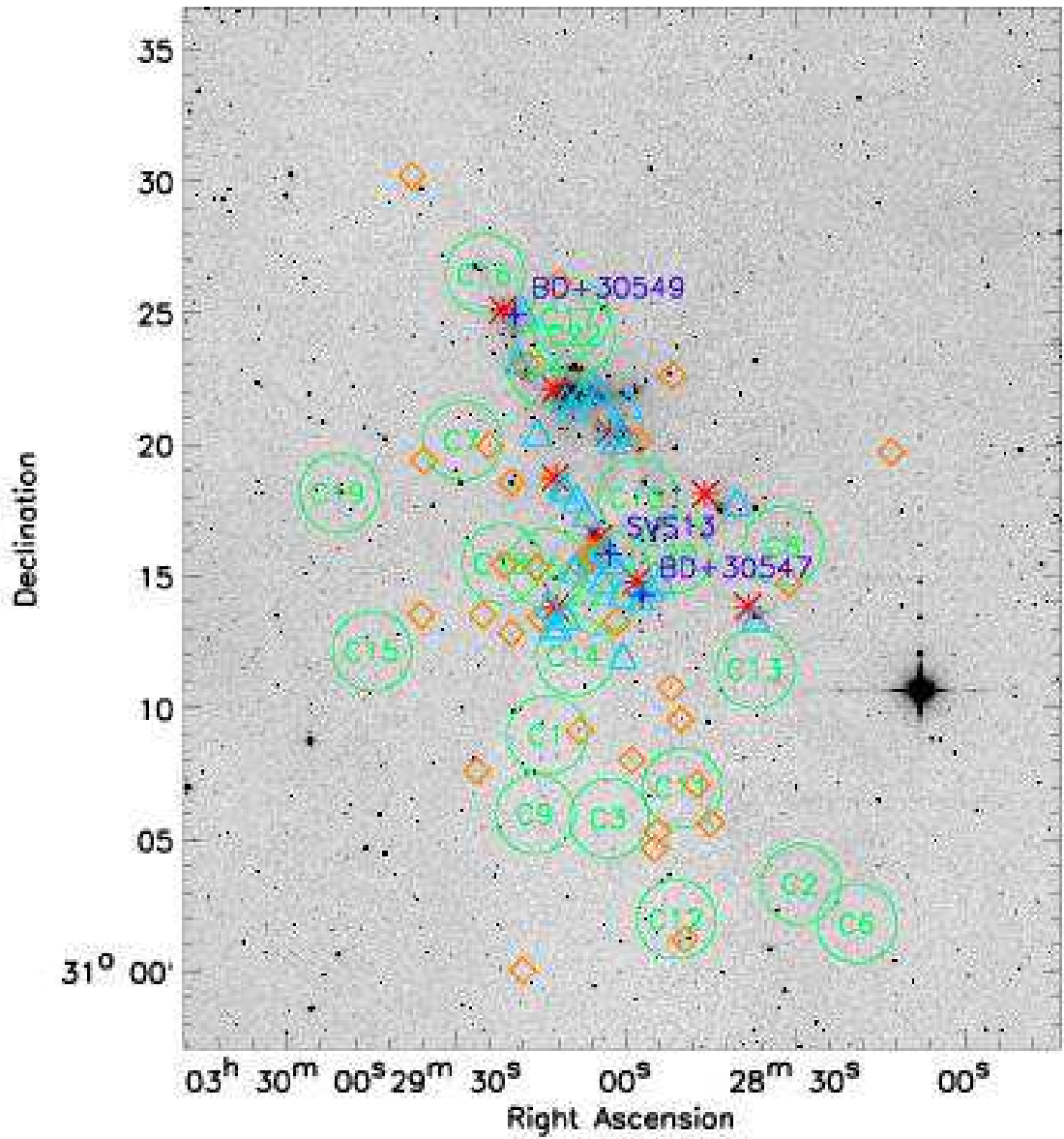}
\figcaption{
Positions of cavities compared to the locations of other sources.
IRAS sources from \citet{jennings} are shown as red stars.  HH objects
from \citet{HHcat} are shown as orange
diamonds. Compact submillimeter sources from \citet{knee}
are shown as blue triangles.
The cavities are shown as green circles of radius 1.5 arcminute.
The grayscale background is the Ks band 2MASS image and shows
T-Tauri cluster stars as well as background stars.
The B stars BD+30547 and BD+30549, and infrared source SVS13
are shown as dark blue crosses.
\label{fig:dolabels}
}
\end{figure*}

In this section we compare the location of the cavities seen in 
individual channel
maps to the positions of sources identified from previous surveys.
A comparison between the location of cavities and
the near-infrared Ks-band 2MASS images showing
T-Tauri stars, young stellar 
clusters and background sources is shown in Figure \ref{fig:dolabels}.
We have plotted our cavity locations along with the positions
of sources identified from infrared imaging surveys,
\citep{aspin,lada96,SVS76}, HH objects \citep{HHcat}
and molecular outflows studied by \citet{knee}.
The positions of the B stars BD+30547 and BD+30549, and infrared source SVS13
are also shown on this figure.

While it was tempting to associated
every cavity with a source, no matter how faint, in practice some
cavities did not contain sources likely to be responsible
for them. For example, cavities on the southern side of the cloud
such as C11, C12 have no HH or 2MASS counterparts.
HH objects and embedded young stellar objects (as seen
from the submillimeter observations of \citealt{sandell}) 
appear to reside within the cloud and are not located in these cavities.
The outflows bright in shocked CO and molecular
hydrogen and those identified as HH objects,  tend to be located
along regions of bright $^{13}$CO emission rather located in cavities.
For example, IRAS 1 is located at a peak in the
$^{13}$CO emission in most channel maps rather than a depression
in any channel map.

The lack of correspondence between the cavities and
molecular outflows, HH objects, submillimeter 
and IRAS sources,
suggests that the cavities are not directly associated with the youngest
stellar objects.  We now consider
the possibility that the cavities are related to older class II sources
which should be visible in the 2MASS Ks band images.
Stars in the stellar clusters evident
in the 2MASS images or the infrared imaging
surveys of \citet{aspin} and \citet{lada96} could
be associated with cavities near the center of the cloud.
The stars studied by \citet{aspin} and \citet{lada96}  (identified from
near-infrared imaging) have ages approximately a 1 million
years old.  Because of the large number
of the stars seen in Ks band on the northern side of the cloud, 
it would be possible to match each cavity in this region 
with a pre-main sequence object.
However the lack of near-infrared objects on the southern and 
eastern regions of the cloud makes it
impossible to associate each cavity with a nearby pre-main
sequence object.
While the lack of correspondence between younger 
sources (embedded with outflows) and cavities
implies that the cavities could be relics of previous activity, the lack
of correspondence between cavities and older sources (see in
the near-infrared images) suggests
that cavities could be created distant from the current location of
the source originally responsible for driving it.

If the cavities are relics, then the stars which could have caused
them may have drifted from their location of birth.
T-Tauri stars 
have been estimated to be approximately a million years old \cite{aspin}.
A star moving at 1 km/s (approximately the velocity dispersion of the cloud)
would be able to move 0.1 pc in $10^5$ years.   This is far enough
that it could have moved across a cavity. 
If the cavities
are relics of outflows, their source stars
could have moved away from the location where they drove outflows.
For example, the embedded class II source IRAS 16316-1540 lies
at the edge of a cavity (associated with RNO91) 
that it could have caused \citet{lee02}.
Cavities could also be relics of long jets (greater than 0.5 pc long)
that have punctured tubular holes through the molecular cloud.
Thus cavities could be formed distant from the source originally
responsible for driving the jet.

In the previous section we pointed out three cavities, (C10, C5 and C19),
which have higher expansion velocities than the others.
Since these three cavities are probably younger, 
it may be easier to find 
candidate objects responsible for causing them.  
C5 and C10 are located in a region that has been covered by most
previous surveys whereas C19 is not, making it difficult
to search for a candidate object responsible for this cavity. 
A variety of types of candidate sources are located in the vicinity of C5.
C10 contains pre-main sequence stars but no IRAS source, CO outflow,
submillimeter source, or HH object (see Figure \ref{fig:dolabels}).  
Since C10 is probably
pointed toward us, we expect the candidate driving source
to be in the vicinity of the cavity.  
This suggests that C10 was driven by one of the nearby
pre-main sequence stars and not a younger source with
an active outflow.

\subsection{Observational summary}

Before we discuss explanations and scenarios accounting
for the cavities we have identified in NGC 1333 in $^{13}$CO we
summarize their observational properties.

1) The molecular cloud of NGC 1333 is full of limb-brightened
shells and cavities which are evident in both channel maps and
position velocity slices in the $^{13}$CO velocity cube.  We have
identified 22 cavities in an approximately $1\times 1$ pc region
on the sky.

2) We can estimate the volume filling factor of these cavities in the cloud.
We have 22 cavities with approximate radii of 0.1pc 
in a volume approximately (1pc)$^3$.
We estimate that approximately 10\% by volume of the molecular cloud is
comprised of cavities.   Lower rim expansion velocity,  lower density contrast
and smaller cavities would be more difficult to detect in the $^{13}$CO velocity
cube. Since we cannot have identified every cavity in this cloud,
we can regard the estimated filling factor as a lower limit.
We conclude that cavities permeate the molecular cloud.

3) Limb-brightening is seen in both position space (making shells)
as well as velocity space, suggesting that cavities can be expanding.

4) Cavities have expansion velocity widths of $\sim 1$ km/s
and spatial widths of 0.1--0.2 pc. The intensity contrast between
that at the rim and that in the center is not high $\sim$2--4,
as estimated from a comparison between the emission intensity 
in the cavity center at the cavity's central velocity
and the emission temperature at its edge.
If the edges of the cavities are turbulent then
the edge would cause emission at a range of velocities.
Consequently the true density contrast between the density
at the rim and that inside the cavity could be larger than our
measured factor.

5) Cavities appear to be non-spherical. They could
be ovate or cylindrical.
Examples of elongated cavities at different orientations are seen
in the velocity cube.

6) We see three cavities with higher velocity $^{13}$CO emission
(above 2 km/s from the cloud mean), suggesting
that some cavities could be younger than others.
There could be an evolutionary sequence.

7)  Some cavities appear to be connected to others.
Tubular cavities are oriented north-south on the southern part
of the cloud and east-west on the eastern side.  This suggests that these
cavities could have been driven by previously active
outflows located near the
center of the cloud where there are million year old star clusters.

8) Cavities are not directly or obviously
associated with outflows previously identified
as HH objects or seen in shocked molecular hydrogen and CO
in Spitzer 4.5$\mu$m images (Porras, private communication).  
Instead these young stellar objects seem to
be associated with dense molecular regions or filaments that are
emitting (rather than deficient) in $^{13}$CO.

Cavity expansion velocities ($\sim 1$km/s) are observed to be
above the sound speed of the cloud, though they are similar
to the size of the cloud's velocity dispersion. 
For gas traveling at the sound
speed of a fraction of a km/s typical
of a cold molecular cloud,  
it would take a few times $10^5$ years for a 0.2pc diameter cavity
to fill in if the cavity were not expanding and assuming that
currently there is no momentum source inside the cavity adding to
the expansion. Turbulent motions in the cloud could lower this
timescale by a factor of a few.
However, it might take an additional few times $10^5$ years
longer to fill in the cavity taking into account the need to
overcome the momentum in the walls implied from the
estimated expansion velocities. Based on the age of
the cluster, estimated from the age of the oldest cluster members or
a few million years, the cavity lifetime is short with respect to the
age of the entire system. This suggests that the cavities
have formed fairly recently, {\it even though they do not seem to
be directly associated with existing outflows}.

Our failure to match cavities with existing outflows suggests that
the cavities identified here are primarily relics and associated with past
outflow activity, as previously suggested by \citet{bally99}.
Whereas existing outflows don't seem to be strongly perturbing the
bulk of the cloud, the cavities we identify here permeate the cloud. It is
possible that we failed to match individual outflows to the
cavities identified here in part because because of the low
angular resolution of the $^{13}$CO velocity cube in comparison to
other studies (e.g., \citealt{arce2004}). The $^{13}$CO velocity
cube used here is only sensitive to cavities larger than $\sim
0.05$ pc since the spatial resolution of these data are 47''
\citep{ridge03}, corresponding to $\sim 0.05$ pc at the distance
of NGC 1333. We expect that cavities with diameter smaller than
0.1pc would not only be harder to detect in the velocity cube
studied here but might also be shorter lived, making them rarer.
In short, smaller cavities could also be associated with molecular 
outflows and HH objects, 
but we may not be able to detect them in the $^{13}$CO cube studied
here if they are too small.

It is possible that the cavities we have identified here are
are result of random turbulence that is excited by energy sources
external to the cluster.  For example, simulations of 
supersonic compressible 
MHD turbulence in a sheering disk can show depressions in 
synthetic channel maps 
(e.g. see Figure 2b of \citealt{pichardo}) 
that mimic the signature of an expanding cavity.
Differentiating between 
organized structures such as shells, cavities
and filaments and those caused
by random turbulent motions is a pervasive problem in the study of the ISM.
For example \citet{elmegreen} at the end of
their recent review of interstellar turbulence ask 
"Why is the power spectrum of density structure
a power law when direct observation shows the ISM to be a collection of
shells, bubbles, comets, spiral wave shocks and other pressurized
structures spanning a wide range of scales?"

To differentiate between structures generated
from sources within the cloud and those arising from turbulent motions
(with energy source assumed to be external to the cloud),
we can attempt to search for differences between the structures
seen in the velocity cube of NGC 1333 and those exhibited
by simulations of turbulent media.  We qualitatively 
compare Figure \ref{fig:gray}
to the synthetic channel maps shown in figure 2b by \citet{pichardo}. 
Depressions in the synthetic
channel maps tend to have redshifted or blueshifted
rims, but not both (as do C11, C13 and C16).  
Depressions in the gas density which span a range of velocities
larger than the velocity dispersion of the cloud (such as C5, C10, and C19)
appear to be absent in the simulation.   This comparison
suggests that conical shaped evacuated regions might be more easily
confused with structures created by turbulence, however ovate or spherical
or higher expansion velocity
cavities are likely to have been driven by internal energy sources.
This comparison suggests that some (but not necessarily
all) of our candidate cavities
have been driven by energy sources internal to the cloud.
But if high velocity cavities are likely to have been driven by internal
sources then we infer that there should also be lower velocity structures
which correspond to older or weaker expanding cavities.
Likewise, if cavities with both redshifted and blue shifted rims are likely
to have been driven by internal sources, then we infer
that there should also be cavities located at the edge of the cloud
which have prominent rims only on one side.
The simulation of \citet{pichardo} illustrates general
properties of supersonic compressible MHD turbulence driven by
energy at large scales, 
however the scale of the simulation (500pc) is larger than
that of our cloud (1pc) and so the simulation is not an ideal physical analogy
to NGC 1333.  It may in future be possible to 
develop better tools to differentiate between the
scenarios for producing structures observed in molecular clouds.

\section{Energetics of Cavities}

To investigate the energetics of driving cavities we consider
wind blown bubble models (e.g., see the review
by \citealt{frank}).  When the bubble is fully radiative we
can consider the momentum-conserving case (e.g., \citealt{koo92a,koo92b}).
We assume that the outflow responsible for the cavity has
ceased, consistent with the interpretation of these cavities
as relics. Therefore, we wish to relate the size and expansion velocity
of the cavity to the total momentum imparted to the ambient cloud
from the outflow.
By dimensional analysis  in the spherical case
\begin{equation}
{P_w \over \rho} \sim R^4 t^{-1}
\label{spherical}
\end{equation}
where $P_w$ is the total momentum output
during the lifetime of the outflow;  $P_w = \dot M_w V_w \tau_w$, where
$\dot M_w$ is the mass flux in the wind at a mean velocity
$V_w$, during a litetime $\tau_w$.
Here $R$ is the radius of the cavity at time $t$ and $\rho$
is the cloud density.
This equation is similar to the Sedov solution for an explosion
and is appropriate if the hydrodynamics can be described
by scale free functions.  
More detailed one dimensional spherical models
(e.g., \citealt{koo92a,koo92b})
support the type of order of magnitude scale-free 
estimate illustrated here.
In more convenient units we find
\begin{equation}
P_w \sim
0.2 M_\odot {\rm km~s^{-1}}
 \left({n(H_2)\over 10^{4} {\rm cm}^{-3}  } \right)
 \left({R \over 0.1  {\rm pc} } \right)^4
 \left({t \over 2 \times 10^5 {\rm yr} }\right)^{-1},
\label{Pw}
\end{equation}
where we have adopted a number density of molecular hydrogen
$n(H_2)\sim 10^4$cm$^{-3}$ typical of the NGC 1333 cloud. Note this
estimate is very sensitive to the radius of the cavity. If the
cavity is one half the size, then the momentum requirements drop by
more than an order of magnitude.  One can also consider the role
of the cavity age in shaping the cloud. If we assume that all
cavities are produced by outflows of identical momentum outputs,
then cavities that differed by a factor of $2$ in size would differ
by a factor of $16$ in age.

For Equation \ref{spherical} we have assumed a spherical model.
We can consider a cylindrical cavity with the scaling
\begin{equation}
{P_{w,\perp} \over \rho l} = R^3 t^{-1}
\label{cylindrical}
\end{equation}
where $P_{w,\perp}$ is the total momentum exerted in the direction
perpendicular to a jet along the jet axis a length $l$, and
$P_{w,\perp}/l$ is the total momentum exerted perpendicular
to the jet axis per unit length. Here $l$ refers to the length of
the cylinder, which
could be half the width of the cloud if the driving source is located
at the center of the cloud and the jet has drilled right through
the cloud.  Using the above scaling relation
we would estimate
\begin{equation}
P_{w,\perp} \sim
0.8 M_\odot {\rm km~s^{-1}}
 \left({n \over 10^{4} {\rm cm}^{-3}  } \right)
 \left({R \over 0.1  {\rm pc} } \right)^3
 \left({l \over 0.4  {\rm pc} } \right)
 \left({t \over 2 \times 10^5 {\rm yr} }\right)^{-1},
\label{Pw_cyl}
\end{equation}
where we have used parameters similar to those in Equation \ref{Pw}.

Both of the above scaling relations
imply that the cavity expands at a velocity
$v_{exp} = {dR \over dt} \sim R/t$.
Different scaling factors (1/3 or 1/4) would be inserted in front
of this relation, depending on the power of $R$ in relations
\ref{spherical} and \ref{cylindrical}.
If the cavities in NGC 1333's molecular cloud 
are $\sim 2 \times 10^5$ years
old we would then predict an expansion velocity close to but
somewhat smaller than that observed.
Consequently for we adopt $t \sim 2 \times 10^5$ years since the outflow
ceased for the age of the 0.1 pc sized cavities in NGC 1333.

We expect that the ratio of momentum exerted perpendicular to the jet
axis to that exerted along the jet axis would depend on a number
of quantities.  For a collimated jet that with a constant
thrust, we expect this ratio would primarily depend upon the Mach number
of the jet. However, this ratio could also
depend on the ratio of momentum originating from a disk wind 
compared to that originating from a collimated jet. 
Young stellar objects probably generate outflows that 
decay with time.   Consequently the bow shock
would slow down as a function of time.  The width of the cavity
compared to its length could increase with time.
The cavity's shape, expansion rate at its edges and density contrast
(inside compared to outside)
should be sensitive to the history of the outflow responsible
for driving it.

A wind-blown cavity model for these cavities can be used
to predict observable cavity properties.
Using Equation \ref{Pw} we can substitute for the expansion velocity
$v_{exp} \sim R/t$ finding
\begin{equation}
{P_w } \sim \rho R^3 v_{exp}
\label{observable}
\end{equation}
where we have placed the observables on the right hand side and left
the total momentum required on the left hand side. 
The observables are the density $\rho$, the cavity size, $R$, and
the expansion velocity $v_{exp}$.
Using Equation \ref{Pw_cyl} and making the same substitution, 
we find 
\begin{equation}
P_{w, \perp } \sim \rho l R^2 v_{exp}.
\label{observable_cyl}
\end{equation}
From these relations we see that a cavity's environment could affect its
dynamic properties. A similar sized cavity in a cloud of higher
density should be older than a similar size one in a lower density
region and so should have a lower expansion velocity. Because it
is easier to detect large fast cavities in a velocity cube, 
we suspect that cavities
will be easiest to identify at the edges of clouds.  At cloud
edges, the
lower densities will allow the cavities to grow to a large size
while maintaining an expansion velocity below the channel
spacing or that of turbulence in the cloud.
This does appear to be the case in NGC 1333, see Figure
\ref{fig:gray}, where it is particularly easy to identify large cavities
near the cloud edge (e.g., C19).

We also expect scaling relations between cavity observables. Older
cavities should be larger, have lower expansion velocities, and
lower density contrasts.  We hope that future studies which
aim to match wind-blown bubble models to young cavities (with
the highest velocities) can 
better constrain the total momentum requirements as well as the
nature of the momentum injection; e.g. wide angle wind or jet, and
momentum injection as a function of time \citep{cunningham05b,thorndike}.

\subsection{Comparison of momentum required to open cavities to that
in observed outflows}

We compare the momentum required to open the cavity (estimated
from Equation \ref{Pw}) to that estimated from young stellar outflows.
Young stellar objects (class 0 and 1 sources) have outflows
lasting $\sim 10^5$ years. Momentum fluxes, estimated from
observations of entrained material, range from 0.1 to $1000
M_\odot$km/s for low and high mass objects respectively 
\citep{richer,fukui}. 
In NGC 1333 itself, \citet{knee} measured a total momentum, due to
about 6 mostly class 0 sources, of $\sim 10 M_\odot$km/s in swept-up high velocity gas (here high velocity means $>$ a few km/s from
the mean). The momentum imparted by the individual outflows ranged
from  $0.2 - 2 M_\odot$km/s. 
We find that these total momentum values are similar to those
estimated from Equations \ref{Pw} and \ref{Pw_cyl} using our scaling relations.
\citet{knee} measured energies of
$\sim 10^{43}$ -- $10^{44}$ erg in each outflow which had typical
velocities of 10km/s based on red-shifted and blue-shifted $^{12}$CO
emission.  

The momentum traced in the $^{12}$CO by \citet{knee} correspond to
sources that are younger than the estimated ages of the cavities.
As the YSOs age we expect that more momentum will be put into the
molecular cloud. The rough correspondence between the momentum
required to open the cavity and that observed in entrained higher
velocity $^{12}$CO suggests that a substantial fraction of the
momentum from outflows observed in $^{12}$CO is required to form
the cavities we have identified in the $^{13}$CO velocity cube. It
may be that only a fraction of the momentum from outflows is
actually ejected from the cloud. {\it Thus we conclude that a
degree of momentum coupling exists between the outflows and the
cloud.} The difference in number of outflows identified in
$^{12}$CO (handful) and the number of cavities we have identified
here (22) implies that cavities can be detected over a timescale
that is a few times longer that that for outflows in $^{12}$CO.

It is likely that there is only a moderately short window in time 
when a cavity would be identified in the $^{13}$CO channel maps.
Cavities with radii smaller than 0.05pc would be difficult to detect
with the 47'' beam of the $^{13}$CO data.
Larger and older cavities,
would be more slowly expanding and would have lower density contrasts.
This would make them more difficult to identify in the channel maps.
Based on Equation \ref{observable} we estimate that
a cavity four times larger could be moving only half as fast.  Thus we expect
a cavity with a 0.4pc diameter would have a shell velocity of only 0.2 km/s
at which time it would become confused with turbulence in the cloud
and so difficult to identify in an individual channel map.
(Note the channel maps here have a comparable velocity 
resolution of 0.133 km/s.)
Above, we have estimated an age of $\sim 2\times 10^5$ years 
for a 0.1 pc radius cavity.  
Equation \ref{observable_cyl} implies that this cavity would have about
one fourth the expansion velocity when it is twice as big, at which time
it would be about eight times older  (using Equation \ref{Pw_cyl}).
From this we estimate that an expanding cavity would be detectable in
the channel maps during a timescale of about a million years.
This estimated timescale compared to that for
outflows (a hundred thousand years) could account for the
difference in number of outflows identified in
$^{12}$CO  and the number of cavities we have identified in $^{13}$CO.

\subsection{Excitation of Turbulence}

Previous studies have suggested that outflows from young stellar
objects are sufficiently energetic or contain sufficient momentum
that they could account for a significant fraction of turbulence
in molecular clouds. Here we have introduced an intermediate
stage, the formation of a cavity which is caused by an extinct or
previously active outflow. In this subsection we compare the total
momentum required to account for the observed cavities to that
present in the turbulent cloud.

The momentum and energy injection rate from outflows in NGC 1333
has been estimated to be $10^{-4}M_\odot {\rm km~s}^{-1} {\rm
yr}^{-1}$ and $0.1 L_\odot$ from the outflows seen in $^{12}$CO
\citep{knee}. In the previous sections we found  
that the momentum
required to form the observed cavities is $\sim 1 M_\odot$km/s,
similar to that of a typical outflow seen in $^{12}$CO by
\citet{knee}.  There are a few times as many cavities (22) as
outflows ($\sim$ 6) seen in $^{12}$CO. The number of observed
cavities is consistent with the longer lifetime we have estimated
for these cavities or a few hundred thousand years. Consequently
the total momentum input into the cloud via the expansion of
cavities is similar to that estimated from the observed outflows
in $^{12}$CO or $\sim 10^{-4} M_\odot$ km~s$^{-1}$ yr$^{-1}$. The
total energy flux consistent with the cavities is also similar to
that estimated  by \citet{knee} or $ \sim 0.1 L_\odot = 4 \times
10^{32}$ erg. 

We note that while we can detect cavities everywhere in
the cloud, the opacity of the $^{12}$CO line makes it difficult
for lower velocity or weak outflows to be detected. For example,
some HH objects are not seen with high velocity $^{12}$CO
counterparts. Also the area of the sky covered by \citet{knee} is
smaller than of our $^{13}$CO velocity cube and didn't include
sources such as IRAS 1 because of this.  It is possible that the
total momentum budget from outflows is somewhat higher than that
estimated  by \citet{knee}.

We now compare the energy flux from outflows
into the cloud to that being dissipated in the cloud.
Since turbulent energy decays in roughly a
crossing time (\citealt{elmegreen} and references their-in),
the energy dissipated by a turbulent medium can be estimated as
\begin{equation}
L_{turb} \sim {M_{cloud} v_{turb}^3 \over l_{eddy}} 
\end{equation}
where 
$l_{eddy}$ is the size of the largest turbulent eddies.
Inserting $M_{cloud} = 520 M_\odot$ for the mass of the cloud (using
that measured by \citealt{ridge03} but correcting for distance),
$v_{turb} \sim 1$km/s and $l_{eddy} \sim 0.3$pc 
we estimate a total energy dissipation rate of
$L_{turb} \sim 10^{33}{\rm erg/s}$.
We insert $l_{eddy} \sim 0.3$ pc corresponding to the approximate size of 
the filaments on the cloud's southern side.
The estimated energy dissipation rate is close
to that available from the outflows seen in $^{12}$CO and
that inferred from the cavities.
This confirms previous
studies that have found that outflows could be responsible for
driving a significant fraction of molecular cloud turbulence
\citep{bally96b,ballyreview,knee,warin}.
However, here we have provided an additional intermediate step between
the driving of the outflow and the excitation of the turbulence, the
slow expansion of multiple cavities which have been opened by
now extinct YSO outflows.

There is an additional test we can carry out.  Previously we estimated
the volume filling factor of the cavities $\sim 10\%$.  We also
have predicted how the cavities evolve in time.  We ask: what
would the cavity expansion velocity be when it reaches a size large enough
to overlap a nearby cavity?  When do random turbulent motions in the cloud
begin to destroy the cavities?   It may not be a coincidence that the cavity
expansion velocities $\sim 1$km/s are similar to the rms
velocity dispersion of the cloud.
From Equation  \ref{observable}
we find that for a given momentum impulse and ambient density, the
expansion velocity is only weakly dependent on the cavity size,
$v_{exp}  \propto R^{1/3}$.
Since the cavity expansion velocity is only weakly dependent on the cavity
size, the expansion velocity would be similar for larger cavities.
The cavities, when they become overlapping, would have expansion
velocities near 1 km/s which is approximately the same as the cloud
rms velocity or velocity dispersion.  Since this velocity is similar
to the random component of the cloud, the cavities will be
destroyed by the turbulence of the cloud, which they
in fact are contributing to.
The high volume filling factor of these cavities, and similar
expansion velocity to the rms velocity dispersion of the cloud
also suggests that they are a significant contributor to the cloud turbulence.



\section{Summary and Discussion}

In this paper we began by studying the structure of the integrated
properties of the $^{13}$CO velocity cube in comparison
to the distribution of young stellar objects seen at near-infrared
wavelengths.  The velocity dispersion of the molecular cloud
is between 1 and 1.6 km/s rms and
varies little across the cloud and does not seem to be closely related to the
distribution of young stellar objects.  The mean velocity
does show structure, with the southern half of the cloud at a mean
velocity that is about 0.3km/s below that of the northern half.
Little evidence for high velocity gas
is seen in $^{13}$CO velocity cube
near the large concentration of young stellar objects 
or outflows identified from other studies.

More careful inspection of the $^{13}$CO velocity cube reveals 
what appear to be numerous cavities and shells.  
These depressions are likely to be real because they are seen in
multiple channels, are present in both channel maps and position
velocity diagrams, and some are also seen in the lower signal
to noise C$^{18}$O velocity cube.  
Rims are limb brightened in individual channel
maps but also tend to
contain brighter gas at higher and lower velocities than the mean.
The features in the position velocity diagrams suggest that the cavities
may not be spherical.  They could be 
expanding ovate or cylindrical structures.
Examples of cavities likely to be at different orientations are seen.
Typical cavity sizes are 0.1--0.2 pc in diameter and have velocity
widths 1--3 km/s.  From the channel maps we estimate
the density contrast between
gas outside and inside the cavity is low, only a factor of 2-4. 
However, the true density contrast could be higher if the edges of the cavities
are turbulent and so contain gas at a range of velocities.
We have identified 22 cavities
in a $1\times 1$ pc region on the sky.
We estimate their volume filling factor to be $\sim 10\%$.
They permeate the molecular cloud.

We find three cavities with higher velocity $^{13}$CO emission
(above 2 km/s from the cloud mean), suggesting
that some cavities could be younger than others.
Some cavities appear to be connected to others.
Tubular cavities are oriented north-south on the southern part
of the cloud and east-west on the eastern side.  These
cavities could have been driven by previously active
outflows located near the center of the cloud where there
are million year old star clusters.

We have had difficulty matching cavities with young stellar objects that
could have been responsible for opening the cavity.
Cavities are not seen near HH objects or objects identified as
outflows from previous surveys.
Neither do the cavities appear
to be clearly associated with million year old pre-main
sequence objects which would be present in the 2MASS Ks-band images.
We estimate that the timescale for a cavity to fill in at the sound speed or at
the random turbulent cloud velocity would not be long, less than
a million years, thus the cavities must have
been produced within the past million years.
If cavities are relics associated with
previous outflow activity, then they could be located
distant from their driving source.  
Also their source
stars could have drifted from their original location.
A star moving 1km/s drifts
out of a 0.1pc sized cavity associated with its formation
in only $10^5$ years.

From scaling laws, we find that expanding bubbles driven by
a jet impulse can match the rough physical properties of the cavities;
i.e., their expansion velocities and sizes.
The cavities are most likely relics associated with
previous outflow activity, as previously suggested by \citet{bally99}
in the case of the Circinus cloud.
We use the dimensional scaling relations
to estimate the age and jet/wind impulse total momentum
required to make a cavity.
This total momentum is about $1 M_\odot$km/s and the estimated
cavity age is a few hundred thousand years.
The age estimate is approximately consistent with
the number of cavities, $\sim 20$, and the number of
active outflows in the cloud seen at higher velocities in $^{12}$CO.

From a comparison between the momentum flux estimated
from active outflows measured from
higher velocity (greater than a few km/s from the mean)
$^{12}$CO by \citet{knee} and the requirements to open
the cavities we estimate that most of
momentum from outflows is fed back into the cloud.
This total mechanical energy is a significant fraction
of that required to power the cloud turbulence.
However, here we have provided an additional intermediate step between
the driving of the outflow and the excitation of the turbulence; the
slow expansion of multiple cavities which have been opened by
now extinct YSO outflows.
The similarity between
the cavity expansion velocity widths and the cloud velocity
dispersion also suggests that the cavities could form an
intermediate step between the outflows and the excitation of
the cloud turbulence.

Our estimates of the momentum requirements of these cavities
are based on rough order of magnitude
scaling arguments.  Our preliminary simulations have been restricted
to jet impulses, rather than jets followed by the slow opening of
a cavity.  Better models of a longer interaction between jets and
winds from young stellar objects with the cloud may allow better constraints
on the properties of the cavities and on how they impact the
evolution of the molecular cloud.
Future studies can explore in more
detail the types of wind blow models which could account
for these cavities, and by fitting observations (channel maps) of
individual cavities better constrain the momentum requirements,
type of injection and outflow evolution 
\citep{thorndike, cunningham05b}.

\acknowledgments
We thank John Bally, Mike Jura, Tom Meageath 
and Alicia Porras for helpful discussions and correspondence.
We thank the Spitzer IRAC GTO team for showing us their Spitzer IRAC images
in advance of publication.
We thank Eve Ostriker and Alyssa Goodman for helpful discussions
on turbulence.
We acknowledge the hospitality of the Aspen Center for Physics during
July 2004.

Support for this work was in part
provided by National Science Foundation
grants AST-9702484, AST-0098442,  
AST 0406799 
and AST-0406823, 
the National Aeronautics and Space Administration
under Grant No.~NNG04GM12G issued through the Origins of Solar
Systems Program, and Grant No.~NAG5-8428.
Support was also provided by
DOE grant DE-FG02-00ER54600, the Laboratory for Laser Energetics and
by the National Science Foundation to the Kavli Institute
for Theoretical Physics under Grant No.~PHY99-07949.

{}

\begin{deluxetable}{lcccc}
\tablecaption{Cavities identified in $^{13}$CO(1-0)}
\tablehead{
\colhead{Name} &
\colhead{RA(J2000)}&
\colhead{DEC(J2000)}&
\colhead{Velocity}&
\colhead{Vel. Range}  
}
\startdata
C1  & 3:29:14 & +31:09:00 &  5.5 & 5.2-7.0 \\
C2  & 3:28:29 & +31:03:20 &  6.0 & ~~~-6.5 \\
C3  & 3:29:03 & +31:05:50 &  6.1 & 5.5-6.5 \\
C4  & 3:29:14 & +31:23:00 &  6.3 & ~~~-6.8 \\
C5  & 3:29:14 & +31:15:20 &  6.3 & 4.5-7.0 \\
C6  & 3:28:19 & +31:01:50 &  6.8 & 6.5-7.5 \\
C7  & 3:29:29 & +31:20:10 &  6.9 & 6.5-8.0 \\
C8  & 3:28:32 & +31:16:10 &  7.0 & ~~~-7.5 \\
C9  & 3:29:16 & +31:06:00 &  7.3 & 6.5-8.0 \\
C10 & 3:28:58 & +31:18:00 &  7.3 & 5.5-10.5\\
C11 & 3:28:50 & +31:07:00 &  7.3 & 6.5-7.8 \\
C12 & 3:28:51 & +31:02:00 &  7.5 & 6.5-7.8 \\
C13 & 3:28:37 & +31:11:30 &  7.6 & 7.0-8.5 \\
C14 & 3:29:09 & +31:12:00 &  7.7 & 7.0-8.3 \\
C15 & 3:29:45 & +31:12:10 &  8.0 & 7.0-9.5 \\
C16 & 3:29:25 & +31:26:30 &  8.0 & 7.5-~~~ \\
C17 & 3:29:10 & +31:24:50 &  8.1 & 7.5-8.5 \\
C18 & 3:29:22 & +31:15:30 &  8.3 & 7.8-9.0 \\
C19 & 3:29:51 & +31:18:10 &  9.1 & 8.5-11.0\\
C20 & 3:28:52 & +31:15:50 &  9.1 & 8.5-~~~ \\
C21 & 3:29:05 & +31:14:20 &  9.1 & 8.0-~~~ \\
C22 & 3:29:09 & +31:24:10 & ~9.3 & 8.5-~~~ 
%
%
%
\enddata
\tablecomments{
Cavities are labeled in $^{13}$CO(1-0) channel maps and
position velocity diagrams and channel
maps shown in Figures \ref{fig:gray},\ref{fig:pv}, and \ref{fig:pvv},
Central cavity RA and DEC are given with EPOCH J2000.
The central velocity is given with respect to
the local standard of rest and is in km/s.
The rightmost column shows the approximate mininum and maximum
velocities (in km/s) of the channels in which the cavity is seen. 
}
\end{deluxetable}

\clearpage

\end{document}